\documentclass{mn2e}

\usepackage{graphicx,amsmath,amssymb,subfigure}

\def\simgt{\mathrel{\lower0.6ex\hbox{$\buildrel {\textstyle >}
 \over {\scriptstyle \sim}$}}}
\def\simlt{\mathrel{\lower0.6ex\hbox{$\buildrel {\textstyle <}
 \over {\scriptstyle \sim}$}}}
\newcommand{\etal}{{et al.}}                   

\newcommand{\Msolar}{\mbox{\,$\rm M_{\odot}$}}        
 
\newcommand{\asec}{^{\prime\prime}}
\newcommand{\fracarcs}{\mbox{\ensuremath{.\!\!^{\prime\prime}}}}%

\newcommand{\gtsim}{\mbox{{\raisebox{-0.4ex}{$\stackrel{>}{{\scriptstyle\sim}}
$}}}}
\newcommand{\ltsim}{\mbox{{\raisebox{-0.4ex}{$\stackrel{<}{{\scriptstyle\sim}}
$}}}}

\begin{document}

\title[Evidence for Cold Accretion onto a Massive Galaxy at High-$z$?]{Evidence for Cold Accretion onto a Massive Galaxy at High Redshift?}

\author[D.J.B.~Smith \& M.J.~Jarvis.]
{Daniel J. B. Smith$^{1}$\thanks{Email: dsmith@astro.ox.ac.uk} \& Matt J.~Jarvis$^{1,2}$\thanks{Email: m.j.jarvis@herts.ac.uk } \\
\footnotesize
$^{1}$Astrophysics, Department of Physics, Keble Road, Oxford, OX1 3RH, UK \\ 
$^{2}$Centre for Astrophysics, Science \& Technology Research Institute, University of Hertfordshire, Hatfield, Herts, AL10 9AB, UK
}

\maketitle

\begin{abstract}

In this letter we report on the discovery of a $z$=2.83 Lyman-$\alpha$
Blob (LAB) found in our wide field narrow-band survey within the
Spitzer First Look Survey region. The blob is extended over at least
95kpc and has a total Lyman-$\alpha$ luminosity of 2.1$\times
10^{44}$ erg s$^{-1}$. It is only the sixth LAB known of this scale
($>$ 50 kpc), and is associated with an embedded continuum source in g',
R, i', K \& 4.5 $\mu$m bands. The LAB's optical spectrum shows
clumpy structures and tantalising hints of a sharp red cut-off and
shear within the Ly-$\alpha$ emission line. Studies of the LAB's
surface brightness profile and of the continuum counterpart's spectral
energy distribution (SED) indicate that the profuse Lyman-$\alpha$
emission is consistent with being powered by cold gas accreting onto a
massive dark matter halo.

\end{abstract}


\section{Introduction}\label{intro}

Recently, much success has been gained in picking out high redshift
galaxies using a narrow-band filter to scan their Lyman $\alpha$
emission in a small redshift ``slice'' of the distant
universe. Steidel \etal~(2000 - hereafter S00), for example,
discovered a `spike' in the redshift distribution of a sample of LBGs
at $z$=3.09 and observed the region with a narrow-band filter tuned to
pick out Lyman $\alpha$ emitting galaxies at that redshift to confirm
the presence of a proto-cluster. They not only found evidence of a
proto-cluster, but also discovered two very extended ($\sim$ 100kpc)
and extremely luminous (L$_{Ly\alpha} \approx 10^{44}$ erg s$^{-1}$)
regions of Lyman $\alpha$ emission with very faint continuum and no
radio detections. They named these objects ``Lyman $\alpha$ Blobs''
(LABs) on account of their almost monochromatic emission and amoebic
appearance.

Perhaps the most mysterious aspect of LABs is the apparent lack of an
obvious source of ionisation to power such profuse Lyman $\alpha$
emission. A variety of models have been suggested to explain this
apparent omission; three of the most plausible explanations are: (i)
the Lyman-$\alpha$ emission comes from a dust-enshrouded, extreme
starburst galaxy with a large scale superwind (e.g. Taniguchi \&
Shioya 2000; Ohyama \etal ~2003; Wilman \etal ~2005), (ii) LABs
contain hidden QSOs (e.g. Haiman \& Rees 2001; Weidinger, M\o ller \&
Fynbo 2004; Weidinger et al. 2005) or (iii) that we are observing the
cooling radiation of a collapsing proto-galaxy inside a dark matter
halo's gravitational potential (the so-called ``cold accretion'' model
- e.g. Haiman, Spaans \& Quataert, 2000, Fardal \etal, 2001, Matsuda
\etal, 2004, Nilsson \etal, 2006) .

Currently only $\sim$ 5 of the most extended ($\gtsim$50kpc scale)
LABs are known (see S00, Francis \etal, 2001, Dey \etal, 2005, Nilsson
\etal, 2006), and it is clear that more are required before we can pin
down the source of ionization with any certainty, or explain the
clustering or evolutionary aspects of LABs within the ever-increasing
compendium of high redshift objects.

In the following sections, we present our wide field survey, give an
account of the discovery of this new LAB, discuss our results \& state
the conclusions we have drawn. Throughout this letter the AB magnitude
system is used (Oke \& Gunn, 1983), and a standard cosmology is
assumed in which $H_{0}$ = 71 km s$^{-1}$, $\Omega_{M}$ = 0.27 and
$\Omega_{\Lambda}$ = 0.73.

\section{Our Survey}\label{oursurvey}

In an attempt to pick out a statistically meaningful sample of $z
\approx 3$ LABs, we have adopted a different approach from other
narrow-band surveys which have tended to concentrate on covering a
small or moderate sky area to large depth (e.g. Matsuda \etal,
2004, Nilsson \etal, 2006 and Saito \etal, 2006). By exploiting the
comparatively large field of view of the Wide Field Camera on the
Isaac Newton Telescope (INT-WFC) our survey has concentrated on
covering a very wide area ($\sim$ 15 sq. degrees) to relatively modest
flux limits.  In this way we can compensate for the necessarily small
co-moving volume probed by a narrow-band filter and as a result we are
sensitive to the brightest, most extended and rarest LABs.

We have observed three areas of sky which allow us to take advantage
of high quality public survey data. The fields we have observed are;
the SWIRE region within the XMM Newton Large Scale Structure survey
field (XMM-LSS) centred on 02$^h$18 -05$^\circ$00 (see Lonsdale \etal,
2003); the Lockman Hole (LH; 10$^{h}$50 +57$^\circ$00; see Lonsdale
\etal, 2003) and the Spitzer First Look Survey (FLS; 17$^h$18
+59$^\circ$30, see Lacy \etal, 2005)

Our survey comprises observations from three different narrow-band
filters, along with complementary observations over the same fields in
Sloan g' and i' bands. The narrow-band filters used (HeII, H$\beta$b,
and [OIII]r) come from the WFC filter set, and are centred on 4686,
4861 \& 5008 \AA, with full-width half maxima of 100, 170, \& 100 \AA,
respectively. For the narrow-band observations discussed in this
letter, we have used the HeII narrow-band filter; the data are found
to be complete to HeII$_{AB} \sim 24.6$, g'$_{AB} \sim 25.0$, and
i'$_{AB} \sim 24.7$ respectively (5$\sigma$), while the seeing in the
narrow-, g'-, and i'-band data are 1.03$\asec$, 1.25$\asec$ and
1.17$\asec$ respectively.

Our data were de-biassed, trimmed, flat-fielded, bad pixel- and
gain-corrected, astrometrized, stacked and photometrically calibrated
using the Cambridge Astronomical Survey Unit WFC pipeline (Irwin \&
Lewis, 2001). General catalogues were generated from matched pointings
using SExtractor (Bertin \& Arnouts, 1996) in double image
mode. Sources of more than 5 adjacent pixels above 1$\sigma$ from the
measured mean in a 4$\asec$ aperture centred on the peak pixel are
source extracted. LAB candidates are selected from our matched
SExtracted g' and HeII catalogues based on their Equivalent Widths
(EWs), which are estimated from their photometry by defining an R
parameter, where

\begin{equation}
R \equiv 10^{0.4\times({\rm g'_{AB}} - {\rm HeII_{AB}})},
\label{Rparam}
\end{equation}

\noindent (from Venemans \etal, 2003). Here g'$_{AB}$ and HeII$_{AB}$
represent the broad- and narrow-band magnitudes of a source. We can
then estimate the observed equivalent widths ($EW_{obs}$) of each
source using

\begin{equation}
EW_{obs} = \frac{R - 1}{\frac{1}{\Delta\lambda_{nb}} - \frac{R}{\Delta\lambda_{g}}},
\label{EW}
\end{equation}

\noindent where $\Delta\lambda_{nb}$ and $\Delta\lambda_{g}$ are the
widths of the narrow- and broad-band filters respectively. LAB
candidates are then selected if they have an $EW_{obs}$ greater than
200\AA\, and $24.1 >$ HeII$_{AB} > 19.5$. It should also be noted that
Hayes \& \"Ostlin (2006) suggest that equivalent widths estimated in
this manner may be a factor of $\sim$2 greater than the true value,
however our follow-up observations do not show this discrepancy (see
section~\ref{results}).

Our narrow-band data are also sensitive to low-redshift line emitting
galaxies. The main contaminant population will be [OII]$_{3727}$ at
$z$ $\approx$ 0.26. Thus, with our selection of line emitting galaxies
at $EW_{obs} > 200$\AA, any [OII] emission line would have to have a
rest-frame equivalent width of $>$ 160 \AA. Hogg \etal (1998) surveyed
375 [OII] emitters within a huge volume spanning 0.0 $\ltsim$ $z$
$\ltsim$ 1.2, and found only two [OII] emitters with $EW_{obs} >$ 100
\AA\, and none with $EW_{obs} >$ 120\AA. A higher fraction of
$EW_{obs} > 120$\AA\, [OII] emitters have been found at higher
redshift (e.g. Ajiki et al. 2006), but this is attributed to evolution
rather than a selection bias at lower redshift. When this is taken
into account, we expect to find $\sim$0.21 $\pm$ 0.46 [OII] emitters
with such high EWs in our coverage of the FLS region. 

Once selected, each candidate must then be checked to confirm that it
is a probable LAB; high proper motion objects, passing satellites,
objects in the haloes of saturated stars and amplifier crosstalk
artefacts must all be eliminated. Thus, each candidate is inspected by
eye to remove these ``false positives'' from the sample. The remaining
candidates are considered robust LAB candidates and await follow-up
spectroscopy.

\section{First Results - A Lyman-$\alpha$ Blob at $z$=2.83}\label{results}

\subsection{The Lyman-$\alpha$ emission}

Figure 1 shows our first spectroscopically confirmed LAB at
17$^h$09$^m$52.3$^s$ +59$^\circ$13$^\prime$ 21$\fracarcs$72 (J2000) in
various wave-bands. It is clearly bright and extended in the HeII
narrow-band filter, while it is comparatively faint and compact in the
g' and i' band images. The Ly-$\alpha$ emitting structure is most
clearly revealed in the continuum-subtracted figure 1(b) and has an
observed equivalent width of $EW_{obs}=657$\AA. The photometric data
for this new LAB at $z$=2.83 are presented in table 1.  We observed
this LAB candidate with the William Herschel Telescope ISIS
spectrograph with the blue and red arms using the R158B and R158R
gratings respectively and with a 2$\asec$ slit width. This results in
a spectral resolution of $\sim$ 15\AA, and wavelength coverage between
$\sim$3500--8300\AA.  The one and two dimensional spectra can be
inspected in figure 2, in which it is clear that the emission is
extended over $12\asec$, corresponding to $\sim$95 kpc at this
redshift.

\begin{table}
\centering
\begin{tabular*}{5.19cm}{|c|c|}
\hline
Photometric Band &  AB Magnitude  \\

\hline
u$^\star$        &  $>$ 26.44                \\
Sloan g'         &  23.97$^{\mbox{ \tiny +0.70}} _{\mbox{ \tiny -0.43}}$   \\
HeII(468.6)      &  22.16$^{\mbox{ \tiny +1.46}} _{\mbox{ \tiny -0.60}}$   \\
R                &  23.81$^{\mbox{ \tiny +0.20}} _{\mbox{ \tiny -0.17}}$   \\
Sloan i'         &  24.07$^{\mbox{ \tiny +1.19}} _{\mbox{ \tiny -0.55}}$   \\
J                &  $>$ 19.74                \\
K                &  21.05$^{\mbox{ \tiny +0.20}} _{\mbox{ \tiny -0.18}}$   \\
4.5 $\mu$m       &  20.71$^{\mbox{ \tiny +0.22}} _{\mbox{ \tiny -0.18}}$   \\
24 $\mu$m        &  $>$ 19.45         \\
610 MHz          &  $>$ 19.14         \\
1.4 GHz          &  $>$ 20.65         \\
\hline
\end{tabular*}
\caption{\label{photometry}Photometric data for the LAB. Magnitudes
are measured in apertures with 4$\asec$ radii, and where they are
limits, they are to 1$\sigma$. The large error in the HeII narrow-band
filter is due to several effects, the faintness of the LAB, the
difficulty in fitting the sky background due to the very extended
nature of the LAB, and the presence of the nearby low-$z$ interloper
to the North and East of the LAB counterpart galaxy. }
\end{table}

\begin{figure*} 
\centering

\subfigure[u$^{\star}$]{\includegraphics[height=1.05in,width=1.05in]{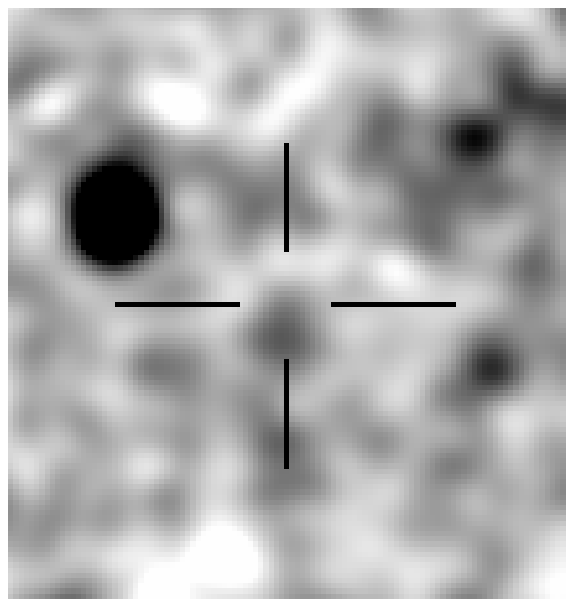} \label{uband1}}
\hspace{0.0cm}
\subfigure[nb-g']{\includegraphics[height=1.05in,width=1.05in]{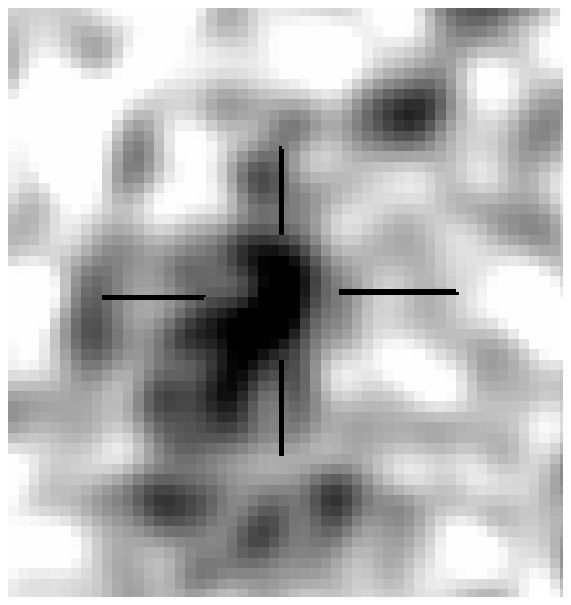} \label{cont_sub}}
\hspace{0.0cm}
\subfigure[HeII nb]{\includegraphics[height=1.05in,width=1.05in]{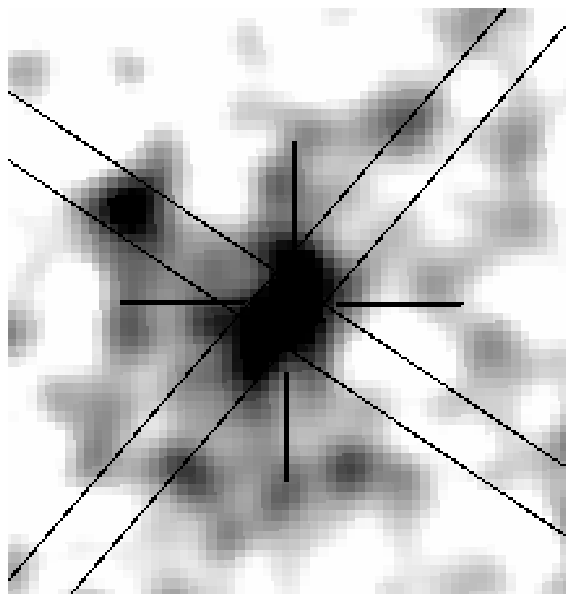} \label{he2band1}}
\hspace{0.0cm}
\subfigure[g']{\includegraphics[height=1.05in,width=1.05in]{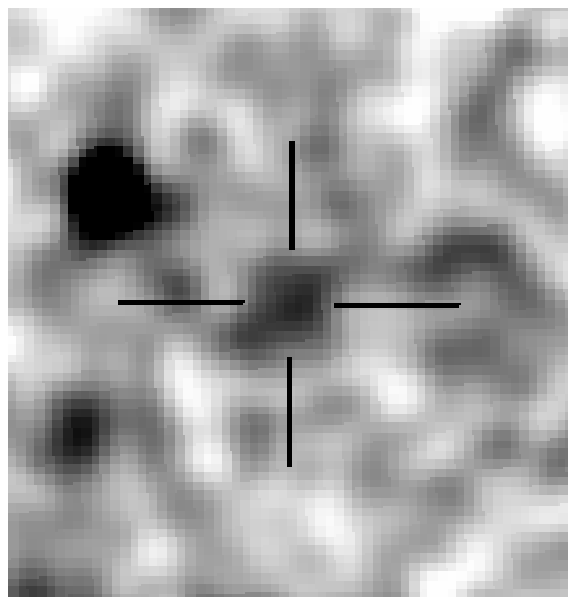}\label{gband1}}
\hspace{0.0cm}
\subfigure[R]{\includegraphics[height=1.05in,width=1.05in]{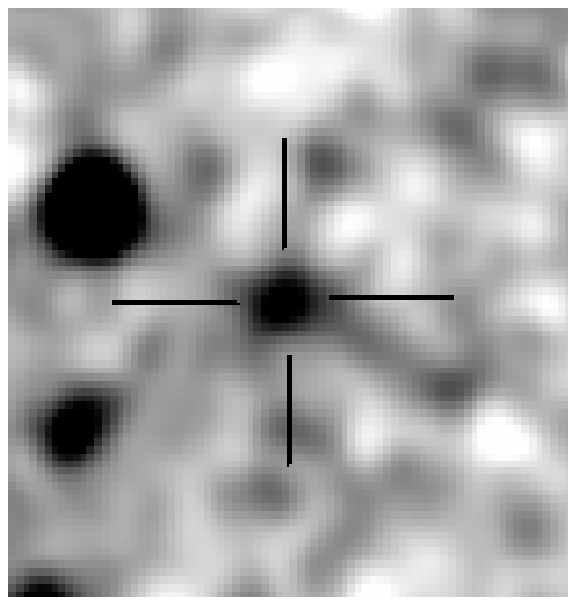} \label{Rband1}}
\\
\vspace{-0.3cm}
\subfigure[i']{\includegraphics[height=1.05in,width=1.05in]{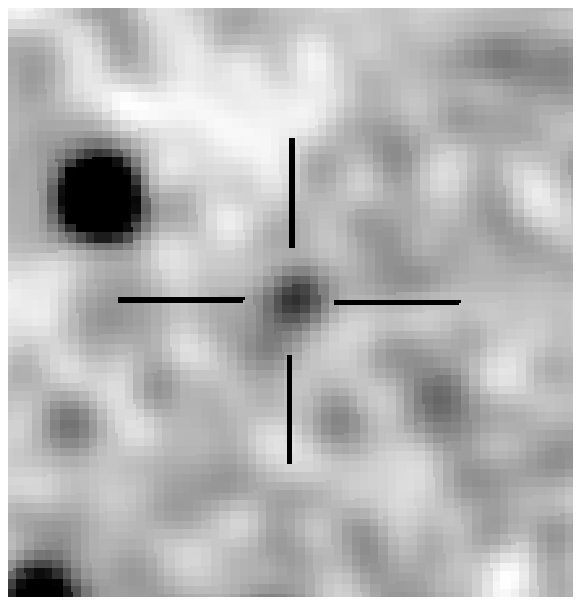} \label{iband1}}
\hspace{0.0cm}
\subfigure[J]{\includegraphics[height=1.05in,width=1.05in]{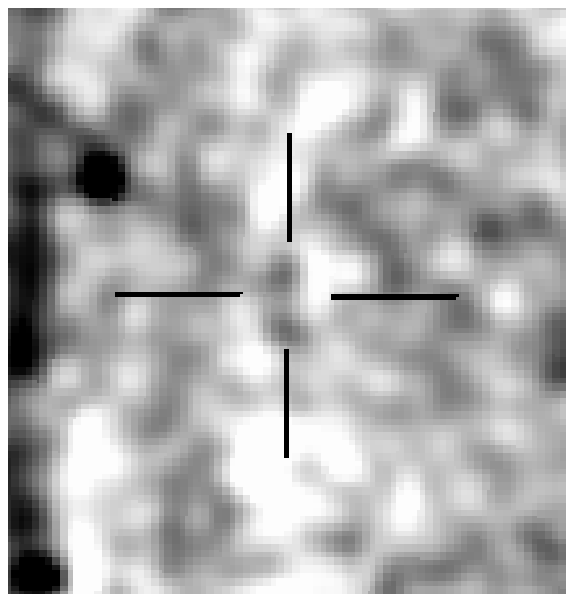} \label{Jband1}}
\hspace{0.0cm}
\subfigure[K]{\includegraphics[height=1.05in,width=1.05in]{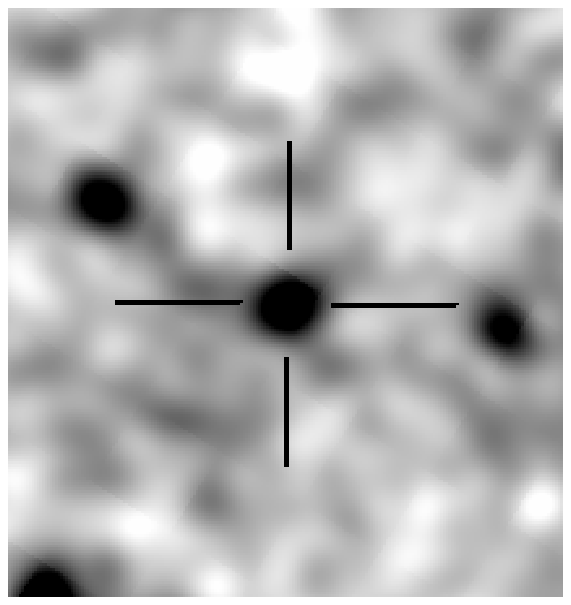} \label{Kband1}}
\hspace{0.0cm}
\subfigure[4.5 $\mu$m]{\includegraphics[height=1.05in,width=1.05in]{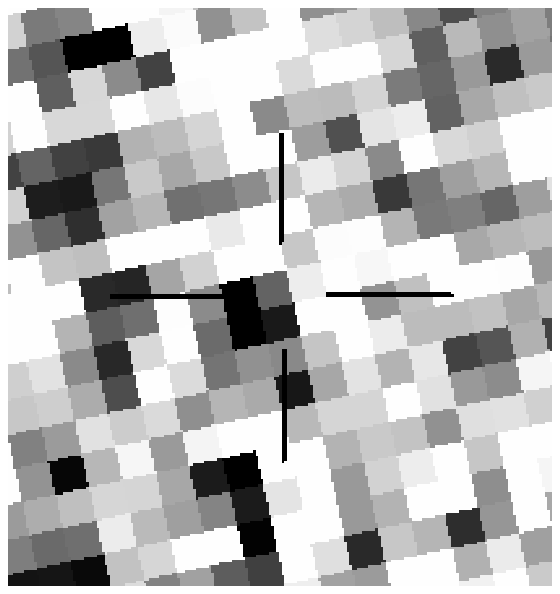} \label{4.5um}}
\hspace{0.0cm}
\subfigure[24 $\mu$m]{\includegraphics[height=1.05in,width=1.05in]{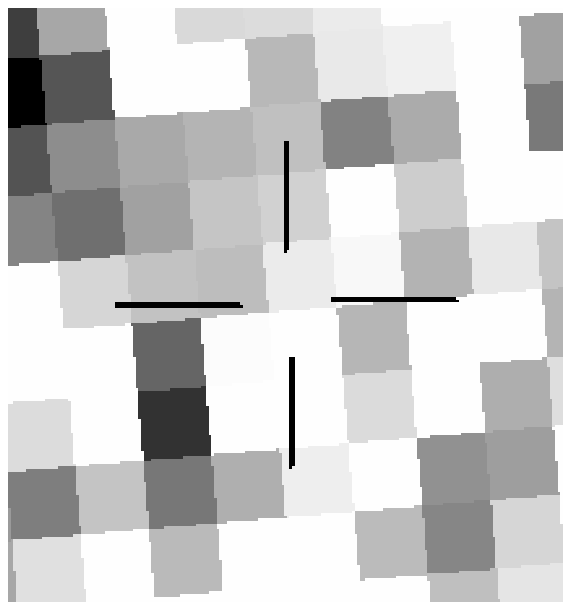} \label{24um}}
\caption{Observations taken in the various bands of the immediate
vicinity of the LAB. The extended Ly-$\alpha$ emission is clearly
visible in the continuum-subtracted and HeII narrow-band images shown
in panes \ref{cont_sub} \& \ref{he2band1}. Figure \ref{he2band1} also
has the slit positions for our follow up spectroscopy shown. Each pane
is $26\times 26$~arcsec$^{2}$ and North is up, with East to the
left. The galaxy to the North and East of the LAB falls within the
slit, and its continuum is apparent in figure \ref{2dspecn2}. It has
been identified as being at $z$$\approx$0.84 due to the presence of
[OII]$_{3727}$, [NeIV]$_{2424}$, MgII$_{2799}$, and [CII]$_{2326}$
emission lines in its spectrum. Figure \ref{Jband1} has variable sky
brightness across it since the LIRIS data have a chip boundary at the
centre of the field of view; it is debatable whether or there is a
detection in this band, and so has been included as an upper limit in
table \ref{photometry}. The u$^{\star}$ band data are from Shim \etal
~(2006), and the R band data come from Fadda \etal ~(2004).}
\label{panes}
\end{figure*}

We identify the line as Ly-$\alpha$ based on its EW, which is too high
to be caused with any reasonable probability by a low-redshift
interloper (see section~\ref{oursurvey}), particularly with the
absence of any other emission lines in our wavelength range (3500 --
8300 \AA).  The FWHM of the line is $\ltsim$ 970 km s$^{-1}$,
unresolved at our spectral resolution, and the total Ly-$\alpha$ flux
across the entire extent of the structure in the ISIS spectrum is
$2.86 \pm 0.07 \times 10^{-16}$ erg s$^{-1}$ cm$^{-2}$, the difference
from the photometric measurement in the narrow-band filter being due
to the geometry of the slit and the 4$\asec$ aperture. Integrating
across the entire extent of the LAB in the continuum-subtracted image,
we estimate the total luminosity of the Ly-$\alpha$ emission to be
$2.1 \times 10^{44}$ erg s$^{-1}$.

The 2D spectra in figure 2 not only show a ``clumpy'' structure, but
the emission line also seems to possess both a cut-off on the red side
of the line (perhaps indicative of a rapid outflow velocity structure
-- see section~\ref{discussion}, below), and hints of a possible shear
which may be interpreted as rotation, although in neither case do we
have the resolution to make a definitive statement; it is clear that
higher resolution spectroscopy is required.
\begin{figure}

\centering

\subfigure{\includegraphics[width=0.85\columnwidth]{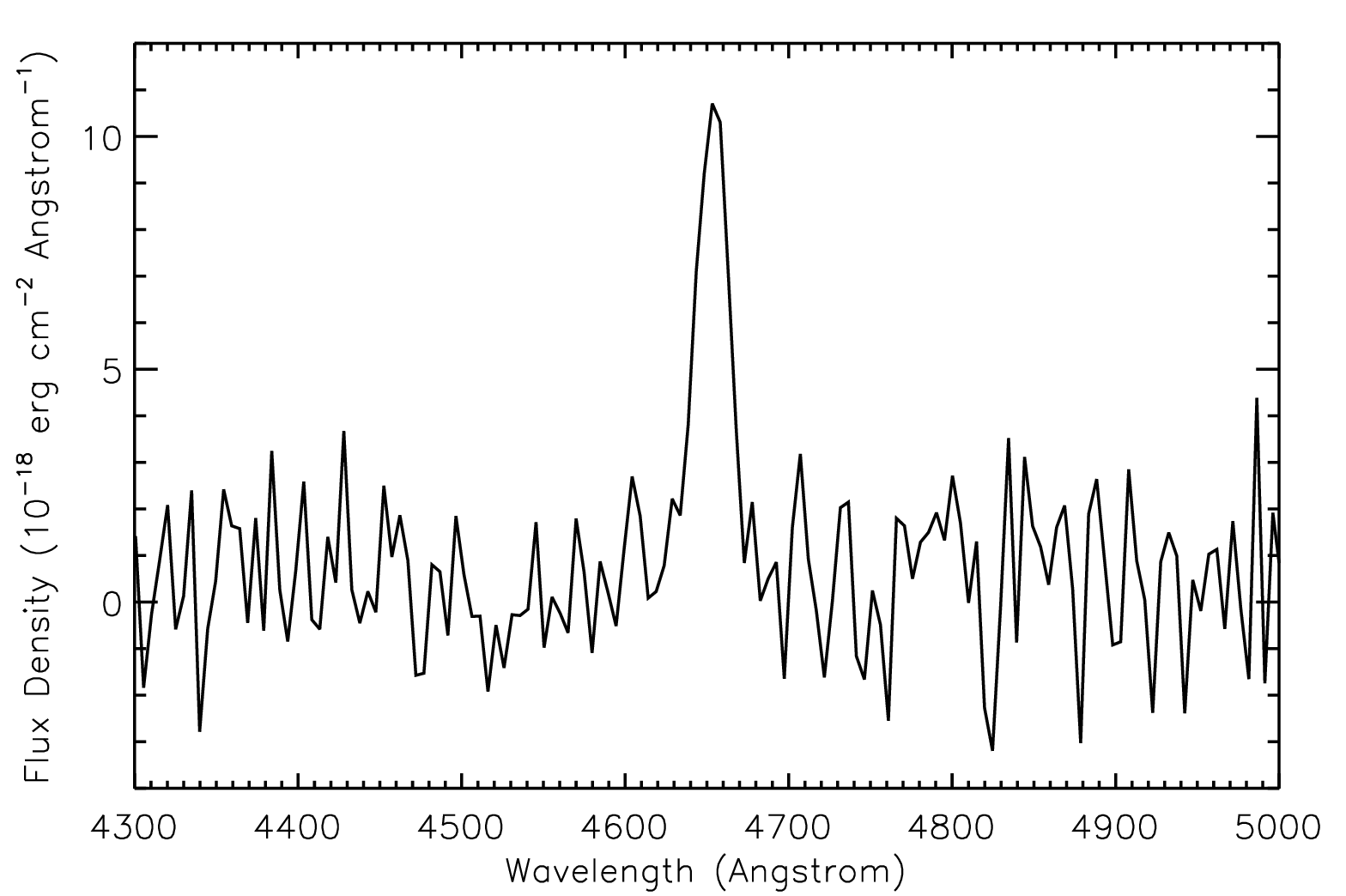}\label{1dspec}}
\subfigure{\includegraphics[width=3.9cm]{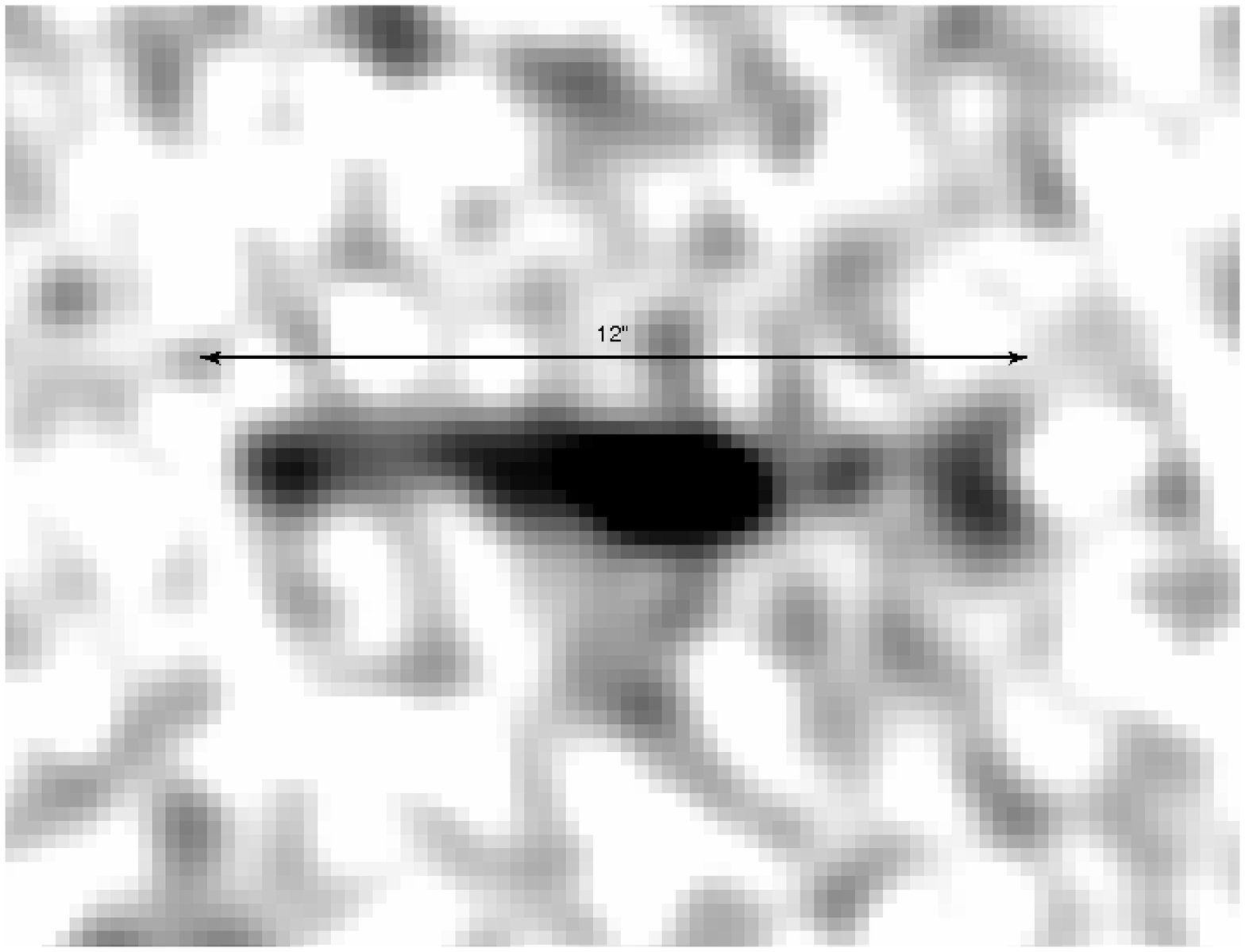} \label{2dspecn1}}
\subfigure{\includegraphics[width=3.9cm]{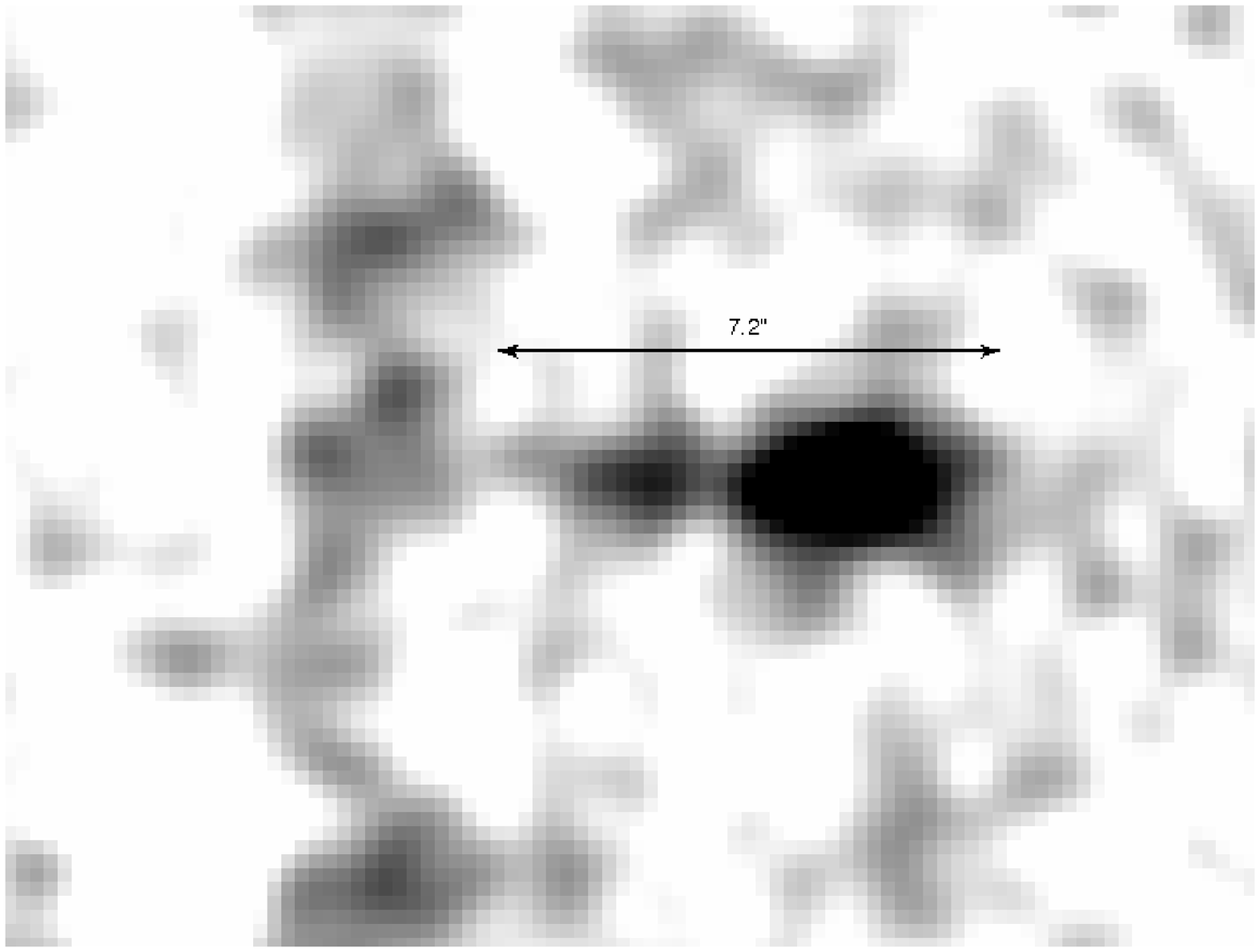} \label{2dspecn2}}

\caption{One- and two-dimensional spectra of the LAB. The 1D spectrum
(top) was extracted over a region 4$\asec$ in the direction SE to NW
[see figure \ref{cont_sub}]. For the 2D spectra (bottom left, and
right), the spatial axis is horizontal and the spectral axis is
vertical. Wavelength increases upwards and the images have been
convolved with a Gaussian kernel with a standard deviation of
0.4$\asec$. The Ly-$\alpha$ emission is extended over at least
$12\asec$, and appears to have a potential cutoff on the red side in
the left hand spectrum along with a slight shear, although the
resolution is low. The right hand spectrum displays continuum emission
to the left of the Ly-$\alpha$ line, which is due to the low-redshift
galaxy to the North-East of the LAB, and not part of the LAB structure
itself. The left hand spectrum was taken at position angle (PA) of
146$^\circ$, while the right hand spectrum was observed at PA =
49$^\circ$. }

\label{2dspec}

\end{figure}

\subsection{The host galaxy}\label{host}

LABs are reminiscent of the large haloes that surround powerful radio
galaxies but with less than 1\% of the associated radio emission
(e.g. Reuland \etal, 2003). Using the FLS-VLA survey at 1.4~GHz of
Condon et al. (2003), we find that there is no detectable radio
emission down to a radio flux-density limit of $S_{\rm 1.4~GHz} \sim$
60$\mu$Jy (3$\sigma$). The LAB is also not detected in 610 MHz data
taken using the Giant Metre-wave Radio Telescope (GMRT) from Garn
\etal ~(2007) to a flux limit of $S_{610\rm MHz}\sim$ 240$\mu$Jy
(3$\sigma$). Thus it is clear that our LAB is significantly fainter
than the radio luminosity of powerful radio sources, with a radio
luminosity of $L_{\rm 1.4~GHz} < 1.6 \times
10^{23}$~W~Hz$^{-1}$~sr$^{-1}$.

At near-infrared wavelengths we obtained service-mode $K-$band data of
the LAB using the UKIRT Fast Track Imager (UFTI; Roche \etal, 2003) on
source for 3780s [Figure~1(h)]. From our observations we find that the
LAB has K$_{AB} = 21.05^{+0.20} _{-0.15}$. We also obtained a 1200s
$J-$band image from the Long-slit Intermediate Resolution Infrared
Spectrograph (LIRIS; Manchado \etal ~1998) on the WHT, in which the
LAB's continuum counterpart is not detected to a 1$\sigma$ magnitude
limit of J$_{\rm AB} >$ 19.74.

The LAB falls just off the edge of the IRAC data in channels 1 \& 3,
but it is covered by channels 2 \& 4 - right on the edge of the
image. Although it looks by eye as though there is a faint detection
in channel four, it is severely affected by reduced signal to noise
near the chip edge and we only show the $4.5~\mu$m data in figure
1. The 24 $\mu$m MIPS data however do cover the region around the LAB
and a thumbnail image is displayed in figure 1(j). There is a faint
source within a 10$\asec$ radius of the i'- and K-band continuum
source, but it is considerably below the completeness limit of
300$\mu$Jy, and is firmly within the noise-dominated regime. It
appears to occupy the same position as a knot of Ly$\alpha$ emission
to the South-East of the centre of the LAB in figure 1(c). If this
24$\mu$m emission is real and associated with the knot of Ly-$\alpha$
in question, it would imply the presence of a dust enshrouded burst of
star formation. However deeper observations at this wavelength are
clearly required to confirm this. It is considered unlikely that the
association is indicative of an obscured AGN since it is offet from
the continuum source in the centre of the LAB by $\sim 50$~kpc at this
redshift, and there is also no other evidence of an AGN anywhere
within the halo.

Furthermore, we fit SED templates to the photometry of the embedded
galaxy using our multi-wavelength data set (figure 3). We use the
Bruzual \& Charlot (2003) model SEDs normalised to the K-band flux to
determine the mass of the embedded galaxy. The K-band flux is used for
the normalisation since it is the least likely to be affected by dust
(the uncertainty on our 4.5$\mu$m point is prohibitively large), and
is thus the most representative measure of the galaxy's old stellar
emission available to us. We use a variety of different populations to
fit synthetic SEDs to our observations; simple stellar populations of
six different metallicities, twenty-one different ages between 10 Myr
and 2.3 Gyr (the age of the Universe at $z$=2.83 is 2.344 Gyr in our
adopted cosmology), and initial mass functions (IMFs) from Chabrier
(2003) and Salpeter (1955) are all used. Our LAB photometric data are
poorly matched by a single simple stellar population (SSP), so linear
combinations of young ($<$ 200 Myr) and old ($\ge$ 500 Myr) SSPs of
varying metallicities and IMFs are used in an attempt to better
approximate the SED.

We find that the best fit to our data points is a combination of a
young (20 Myr) and an old (1.0 Gyr) SSP with a Chabrier IMF and solar
metallicity (Z $= 0.02$). In particular, dust extinction is not
required to give a good quality fit to the data, consistent with the
absence of any significant 24$\mu$m detection. With this SED fit, we
calculate a stellar mass for this LAB of M$_{stellar} = 2.79 \pm 0.38
\times 10^{11}$~\Msolar, consistent with the most massive galaxies
observed at high redshifts such as the hosts of powerful active
galactic nuclei (e.g. Jarvis et al. 2001; McLure et al. 2006), and
find that $\sim$20\% of the LAB counterpart's stellar luminosity comes
from the young starburst population, with the remaining $\sim$80\%
arising from the old population.

We now consider the possibility that the counterpart galaxy within the
LAB may be supplying the source of ionization for the LAB. We estimate
from the model SED that the counterpart galaxy has a bolometric
luminosity (blue-ward of the rest-frame Lyman limit, as would be
expected for case B recombination) of $\sim$2.4$ \times 10^{42}$ erg
s$^{-1}$. Thus we conclude that at most only $\ltsim$1\% of
the LAB's Lyman $\alpha$ emission may be powered by the counterpart
galaxy.

\section{Discussion}\label{discussion}

In this section we attempt to distinguish between the three ionization
models for LABs discussed in section \ref{intro}. 

\begin{figure}
\centering
\includegraphics[width=0.85\columnwidth]{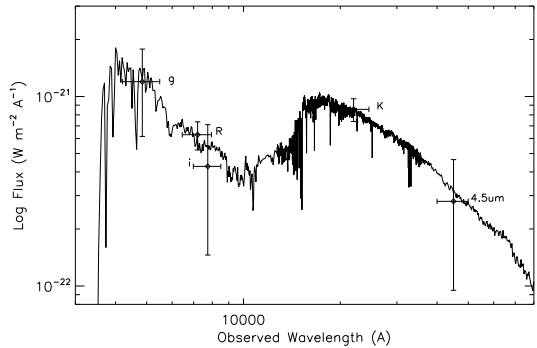}
\caption{The SED of the galaxy immersed within the LAB (error bars),
overlaid with the best fit redshifted simple stellar population models
of Bruzual \& Charlot (2003). The best fit consists of a combination
of two stellar populations aged 30 Myr and 1.0 Gyr, both with
metallicity Z = 0.02. $\sim$20\% of the LAB's bolometric luminosity
comes from the young population, with the remaining $\sim$80\% coming
from the older population.}
\label{SED}
\end{figure}

In the absence of detections in both the 24$\mu$m and 1.4 GHz bands,
the obscured QSO ionization model is highly unlikely. The flux limits
in these bands essentially rule out the presence of a powerful AGN
(both obscured and unobscured). Moreover, the absence of any
high-ionisation lines (such as CIV or CIII) indicative of an AGN,
coupled with an optical\slash infrared SED (figure \ref{SED}) that is
approximated very well by two simple stellar populations, give no
indication of the presence of an AGN at all.

To estimate whether the central galaxy is a massive starburst galaxy
we estimate its star-formation rate (SFR) from three different
diagnostics. These are based on the ultra-violet, radio and
Ly-$\alpha$ luminosities (see Brocklehurst 1971, Kennicutt 1998, and
Yun, Reddy \& Condon 2001), where
\begin{align}\label{sfreqn}
{\rm SFR} (\Msolar ~yr^{-1}) &= 1.4 \times 10^{-28} ~L_{UV} {\rm(erg~s^{-1} Hz^{-1})} \nonumber \\
                             &= 5.9 \pm 1.8 \times 10^{-29} ~L_{1.4 GHz} {\rm(erg ~s^{-1} Hz^{-1})} \nonumber \\
                             &= 9.1 \times 10^{-43} ~L_{Ly\alpha} {\rm (erg~s^{-1})}.
\end{align}

\noindent For our LAB we find using the UV continuum luminosity
(estimated from the SED fit) that the SFR of the embedded galaxy is
23.9 \Msolar yr$^{-1}$. The radio data only provides a large upper
limit ($\ltsim 1200$\Msolar) so is not considered further. The SFR
based on the Ly-$\alpha$ emission of the counterpart galaxy is more
complicated, since it is difficult to separate the Ly-$\alpha$
emission of the galaxy from that of the halo. Thus, to estimate the
SFR of the host galaxy we use the Ly$\alpha$ luminosity measured in
the central 4~arcsec which gives SFR $\approx 18$~\Msolar, consistent
with our UV-based estimate. These optical SFR estimators assume the
absence of dust; any extinction would imply that these estimates
represent a lower limit.  However, we find no evidence for large dust
masses using both the properties of our SED fit and the absence of any
detectable reprocessed emission at 24$\mu$m.

These results are highly significant since Fardal \etal ~(2001), using
hydrodynamic simulations of the cold accretion model, predict that
LABs should have moderate star-forming galaxies at their
centres. Although they predict slightly larger SFRs of
60--600\Msolar~yr$^{-1}$, our values are within the right order of
magnitude (particularly given the uncertainty in both the models and
in the relations given in equations~\ref{sfreqn}).  Furthermore, the
cold accretion models from Dijkstra \etal ~(2006) also predict a sharp
cut-off on the red side of the Ly-$\alpha$ line of the type hinted at
by the red edge of our emission line in figure 3(a).

On the other hand, the superwind model of Matsuda \etal ~(2004),
estimates a SFR of 600 \Msolar yr$^{-1}$ for a 100kpc LAB, and draws a
parallel between this and the submillimeter detection (and thus high
SFR $\sim$ 1000 \Msolar yr$^{-1}$) of LAB1 from S00 (Chapman \etal,
2001). Our SFRs are considerably more quiescent; they may be able to
cause some outflow, but it is probably unlikely that a starburst of
this magnitude could cause superwinds of the types suggested by
Taniguchi \& Shioya (2000), Ohyama \etal (2003) or Matsuda \etal
(2004).

\begin{figure}
\centering
\includegraphics[width=0.85\columnwidth]{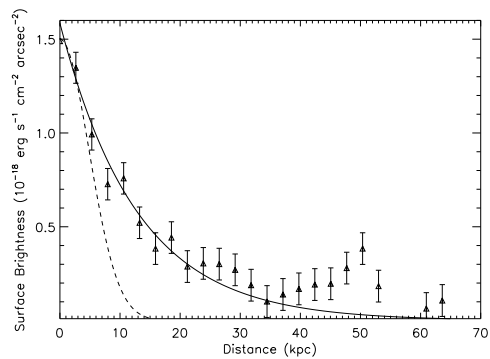}
\caption{The continuum-subtracted surface brightness profile of the
  LAB from our narrow- and broad-band data. The flux-density is
  sky-subtracted emission line flux-density calculated in circular
  apertures, and the PSF of the image is shown with the dashed
  line. The solid line represents ``model 2'' from Dijkstra et al. (2006).}
\label{profile}
\end{figure}

To further evaluate the likelihood of this LAB being a collapsing
cloud of neutral hydrogen in a dark matter halo, we compare the
Ly-$\alpha$ surface brightness profile (SBP) of the new LAB (figure 4)
with that predicted by Dijkstra \etal ~(2006). We find that there is a
very good fit between our SBP and extended cold-accretion ``Model 2''
from Dijkstra's paper, with a total mass of 5.2 $\times 10^{11}$
\Msolar. The biggest deviation from Dijkstra's model is an apparent
increase in the continuum-subtracted surface brightness around r
$\sim$ 50kpc. This is clearly not due to the nearby galaxy to the
North-East of the blob centre as demonstrated in figure 1(b); it is
rather a manifestation of clumpy sub-structure within the LAB. This is
not predicted by the cold-accretion model of Dijkstra (2006), which
assumes the gas within a LAB to be spherically symmetric. Future
higher-resolution spectroscopy over the whole halo would help to
distinguish between a starburst driven superwind and the
cold-accretion model, however the current evidence favours the cold
accretion model.

\section{Conclusions}\label{conclusions}

\begin{itemize}
\item We have presented our wide-field narrow-band survey, and
demonstrated its ability to pick out LABs at high redshift. Over the
whole survey we expect to find a significant number of LABs to
determine their nature fully.

\item We have shown the first confirmed LAB from our survey as being
extended over $\sim$95kpc and having a Lyman $\alpha$ luminosity of
2.1$\times 10^{44}$ erg s$^{-1}$.

\item Our LAB is best described by the cold accretion ionization models of
Fardal (2001) and Dijkstra (2006), although we cannot completely rule
out the superwind model. It is highly unlikely that our LAB contains
an AGN component. Further studies are required to strengthen our
confidence in these assertions and also to gain better mass estimates
for this LAB.

\end{itemize}

\section*{ACKNOWLEDGEMENTS} 
{\scriptsize The authors would like to thank Eduardo
Gonz\'alez-Solares for his assistance with the CASU WFC pipeline,
Richard Wilman, Garret Cotter \& Alejo Mart\'inez-Sansigre for
valuable discussions, and Hyunjin Shim for proprietary u$^{\star}$
band data. DS thanks the UK PPARC for a studentship. The INT and WHT
are operated on the island of La Palma by the Isaac Newton Group in
the Spanish Observatorio del Roque de los Muchachos of the Instituto
de Astrofisica de Canarias. The National Optical Astronomy Observatory
(NOAO) is operated by the Association of Universities for Research in
Astronomy (AURA), Inc. under cooperative agreement with the National
Science Foundation. The United Kingdom Infrared Telescope is operated
by the Joint Astronomy Centre on behalf of the U.K. Particle Physics
and Astronomy Research Council and some of the data reported here were
obtained as part of the UKIRT Service Programme.}

{} 


\begin{thebibliography}{} 

    \bibitem{Ajiki06} Ajiki M., \etal, 2006, PASJ, {\bf 58}, 113A
    \bibitem{SExtractor} Bertin E., \& Arnouts S., 1996, A\&AS, {\bf 117}, 393B
    \bibitem{Brocklehurst71} Brocklehurst M., 1971, MNRAS, {\bf 153} 471B
    \bibitem{Bruzual03} Bruzual G., \& Charlot S., 2003, MNRAS, {\bf 344}, 1000B
    \bibitem{Chabrier03} Chabrier G., 2003, PASP, {\bf 115}, 763C
    \bibitem{Chapman01} Chapman S.C., Richards E.A., Lewis G.F., Wilson G., \&Barger A.J., 2001, ApJ, {\bf 548}, L17 
    \bibitem{Condon03} Condon J.J., Cotton W.D., Yin Q.F., Shupe D.L., Storrie-Lombardi L. J., Helou G., Soifer B.T., Werner M.W., 2003, AJ, {\bf 125}. 2411
    \bibitem{Dey05} Dey \etal, 2005, ApJ, {\bf 629}, 654D 
    \bibitem{Dijkstra06} Dijkstra M., Haiman Z., Spaans M., 2006, ApJ, {\bf 649}, 14D
    \bibitem{Fadda04} Fadda D., Jannuzi B., Ford A., \& Storrie-Lombardi, L.J., 2004, AJ, {\bf 128}, 1
    \bibitem{Fardal01} Fardal M.A., Katz N., Gardner J.P., Hernquist, L., Weinberg D.H., \& Romeel D., 2001, ApJ, {\bf 562}, 605 
    \bibitem{Francis01} Francis P.J., \etal, 2001, ApJ, {\bf 554}, 1001F
    \bibitem{Haiman00} Haiman Z., Spaans M. \& Quataert E., 2000, ApJ, {\bf 537}, L5 
    \bibitem{Haiman01} Haiman Z., \& Rees M.J., 2001, ApJ, {\bf 553}, 545H
    \bibitem{Hayes06} Hayes M., \& \"Ostlin G., 2006, A\&A, {\bf 460}, 681H  
    \bibitem{Hogg98} Hogg D.W., Cohen J.G., Blandford R., \& Pahre M.A.,
    1998, ApJ, {\bf 504}, 622
    \bibitem{Irwin01} Irwin M., \& Lewis J., 2001, NewAR, {\bf 45}, 1051
    \bibitem{Jarvis01} Jarvis M.J., Rawlings S., Eales S.A., Blundell K.M., Bunker A.J., Croft S., McLure R.J., Willott C.J., 2001, MNRAS, {\bf 326}, 1585.
    \bibitem{Kennicutt98} Kennicutt R.C., 1998, ARA\&A, {\bf 36}, 189K
    \bibitem{Lacy05} Lacy M., \etal, 2005, ApJS, {\bf 161}, 41.
    \bibitem{Lonsdale03} Lonsdale, C. J. \etal, 2003, PASP, {\bf 115}, 897
    \bibitem{Matsuda04} Matsuda Y., \etal, 2004, ApJ, {\bf 128}, 569
    \bibitem{Macdonald04} MacDonald E.C., \etal, 2004, MNRAS, {\bf 352}, 1255
    \bibitem{Manchado98} Manchado A. \etal, 1998, SPIE, {\bf 3354}, 448M
    \bibitem{McLure06} McLure R.J., Jarvis M.J., Targett T.A., Dunlop J.S., Best P.N., 2006, MNRAS, {\bf 368}, 1395
    \bibitem{Nilsson06} Nilsson K.K., Fynbo J.P.U., M{\o}ller P., Sommer-Larsen J., Ledoux C., 2006, A\&A, {\bf 452}, 23N
    \bibitem{Ohyama03} Ohyama Y., \etal, 2003, ApJ, {\bf 591}, 9O 
    \bibitem{Oke83} Oke, J.B. \& Gunn J.E., 1983, ApJ, {\bf 266}, 713O
    \bibitem{Reuland03} Reuland M., \etal, 2003, ApJ, {\bf 592}, 755
    \bibitem{Roche03} Roche P., \etal, 2003, SPIE, {\bf 4841}, 901R
    \bibitem{Saito06} Saito T., Shimaksaku K., Okamura S., Ouchi M., Akiyama M., Yoshida M., 2006, ApJ, {\bf 648}, 54
    \bibitem{Salpeter55} Salpeter E.E., 1955, ApJ, {\bf 121}, 161S
    \bibitem{Shim06} Shim H., Im M., Pak S., Choi P., Fadda D., Helou G., \& Storrie-Lombardi L., 2006, ApJ, {\bf 164}, 435
    \bibitem{Steidel00} Steidel C.C., Adelberger K.L., Shapley A.E., Pettini M., Dickinson M., Giavalisco M., 2000, ApJ, {\bf 532}, 170 
    \bibitem{Taniguchi00} Taniguchi Y. \& Shioya Y., 2000, ApJ, {\bf 532}, L13 
    \bibitem{Venemans03} Venemans B.P., Kurk J.D., Miley G.K., R\"ottgering H.J.A., 2003, NewAR, {\bf 47}, 353V 
    \bibitem{Weidinger04} Weidinger M., M\o ller P., Fynbo J.P.U., 2004, Nature {\bf 430} 999W
    \bibitem{Weidinger05} Weidinger M., M\o ller P., Fynbo J.P.U., Thomsen B., 2005, A\&A, {\bf 436} 825W
    \bibitem{Wilman05} Wilman R.J., Gerssen J., Bower R.G., Morris S.L., Bacon R., de Zeeuw P.T., Davies R.L., 2005, Nature, {\bf 436}, 227



\end{thebibliography}
\end{document}